\documentclass[sigconf, review]{acmart}
\copyrightyear{2024}
\acmYear{2024}
\setcopyright{acmlicensed}
\acmConference[CIKM '24] {Proceedings of the 33rd ACM International Conference on Information and Knowledge Management}{October 21--25, 2024}{Boise, ID, USA.}
\acmBooktitle{Proceedings of the 33rd ACM International Conference on Information and Knowledge Management (CIKM '24), October 21--25, 2024, Boise, ID, USA}
\acmISBN{979-8-4007-0436-9/24/10}
\acmDOI{10.1145/3627673.3680077}

\usepackage{algorithm}
\usepackage{algorithmic}
\usepackage{amsmath}
\usepackage{threeparttable}
\usepackage{makecell}
\usepackage{multirow}
\usepackage{diagbox}
\usepackage{graphicx}
\usepackage{booktabs}
\usepackage{siunitx}
\usepackage[section]{placeins}
\usepackage{enumitem}

\AtBeginDocument{%
  \providecommand\BibTeX{{%
    \normalfont B\kern-0.5em{\scshape i\kern-0.25em b}\kern-0.8em\TeX}}}

\setcopyright{none}
\renewcommand\footnotetextcopyrightpermission[1]{}
\begin{document}
\title{Deep Uncertainty-Based Explore for Index Construction and Retrieval in Recommendation System}

\author{Xin Jiang}
\affiliation{
 \institution{Shopee Discovery Ads}
 \city{Beijing}
 \country{China}
 }
\email{andy.jiang@shopee.com}

\author{Kaiqiang Wang}
\affiliation{
 \institution{Shopee Discovery Ads}
 \city{Beijing}
 \country{China}
 }
\email{kaiqiang.wang@shopee.com}

\author{Yinlong Wang}
\affiliation{
 \institution{Shopee Discovery Ads}
 \city{Beijing}
 \country{China}
 }
\email{yinlong.wang@shopee.com}

\author{Fengchang Lv}
\affiliation{
 \institution{Shopee Discovery Ads}
 \city{Beijing}
 \country{China}
 }
\email{fengchang.lv@shopee.com}

\author{Taiyang Peng}
\affiliation{
 \institution{Shopee Discovery Ads}
 \city{Beijing}
 \country{China}
 }
\email{taiyang.peng@shopee.com}

\author{Shuai Yang}
\affiliation{
 \institution{Shopee Discovery Ads}
 \city{Beijing}
 \country{China}
 }
\email{lucas.yang@shopee.com}

\author{Xianteng Wu}
\affiliation{
 \institution{Shopee Discovery Ads}
 \city{Beijing}
 \country{China}
 }
\email{xianteng.wu@shopee.com}

\author{Pengye Zhang}
\affiliation{
 \institution{Shopee Discovery Ads}
 \city{Beijing}
 \country{China}
 }
\email{pengye.zhang@shopee.com}

\author{Shuo Yuan}
\affiliation{
 \institution{Shopee Discovery Ads}
 \city{Beijing}
 \country{China}
 }
\email{shuoyuan.ys@shopee.com}

\author{Yifan Zeng}
\affiliation{
 \institution{Shopee Discovery Ads}
 \city{Beijing}
 \country{China}
 }
\email{yifan.zeng@shopee.com}

\renewcommand{\shortauthors}{Xin Jiang, et al.}

\begin{abstract}
In recommendation systems, the relevance and novelty of the final results are selected through a cascade system of Matching -> Ranking -> Strategy. The matching model serves as the starting point of the pipeline and determines the upper bound of the subsequent stages. Balancing the relevance and novelty of matching results is a crucial step in the design and optimization of recommendation systems, contributing significantly to improving recommendation quality. However, the typical matching algorithms have not simultaneously addressed the relevance and novelty perfectly. One main reason is that deep matching algorithms exhibit significant uncertainty when estimating items in the long tail (e.g., due to insufficient training samples) items. 

The uncertainty not only affects the training of the models but also influences the confidence in the index construction and beam search retrieval process of these models.

This paper proposes the UICR (\textbf{U}ncertainty-based explore for \textbf{I}ndex \textbf{C}onstruction and \textbf{R}etrieval) algorithm, which introduces the concept of uncertainty modeling in the matching stage and achieves multi-task modeling of model uncertainty and index uncertainty. The final matching results are obtained by combining the relevance score and uncertainty score infered by the model. Experimental results demonstrate that the UICR improves novelty without sacrificing relevance on real-world industrial productive environments and multiple open-source datasets. Remarkably, online A/B test results of display advertising in Shopee demonstrates the effectiveness of the proposed algorithm.
\end{abstract}


\maketitle

\section{Introduction}
Recommendation systems adopt a multi-stage cascade architecture \cite{fayyaz2020recommendation,khanal2020systematic}, primarily comprising matching, ranking and strategy. During the matching stage\cite{xu2018deep}, thousands of candidates are retrieved from an extensive corpus. In the subsequent ranking and strategy stages, these retrieved candidates are prioritized based on user preferences and business rules. One of the core objectives of recommendation systems is to deliver recommendations that are not only relevant to users but also exhibit novelty \cite{zhao2016much,fu2023modeling,lo2021ppnw,nugroho5user}. Novelty typically refers to items or content recommended to users that they have not encountered in their historical behavior or exposure. It is important to note that the set of items considered novel varies for different users. Therefore, at the forefront of the system pipeline, in the matching stage, it is crucial to balance the relevance and the novelty.

In industrial-scale recommendation systems, the matching service based on deep models, such as \cite{covington2016deep,chen2022approximate,huang2013learning}, often needs to retrieve thousands of items from millions to billions of candidate items within tens of milliseconds for subsequent scoring in the downstream pipeline. To accelerate the matching process, recommendation systems often employ a data structure called an index, such as the HNSW\cite{malkov2018efficient} algorithm, to assist the matching model in storing and retrieving candidate items relevant to user interests.

During the training process of deep matching models, due to the lack of sufficient training samples for long-tail items \cite{li2023stan,gong2023attention,zhang2023deep}, traditional point estimation training methods fail to adequately train the embeddings for these items. This leads to two problems: 1) Under traditional point estimation modeling, these models exhibit significant uncertainty in predicting long-tail items (e.g., due to insufficient sample size\cite{samuel2021distributional,menon2020long,wang2020long}). 2) The index, constructed by calculation upon insufficient trained embeddings, is lack uncertainty information, thus unable to provide high relevance and low uncertainty items set to the deep matching model, resulting in a loss of relevance in the final matching results. Uncertainty refers to the degree of uncertainty in the recommendation system's predictions or estimates of user interests, preferences, or behavior. It reflects the system's understanding of users and the reliability of the recommendation results. Therefore we introduce variance and transform the problem from a point estimation of the user interest to a distribution estimate, use variance to characterize the magnitude of uncertainty. At the same time, these matching algorithms focus more on improving relevance and pay insufficient attention to novelty metrics. In some cases, the improvement in relevance metrics comes at the expense of novelty metrics, which is not conducive to the long-term ecological development of recommendation systems.

To address the aforementioned issues, this paper proposes the UICR method. UICR comprises three main components: Uncertainty-based Index (UN-Index), Uncertainty-based Retrieval (UN-Retrieval), and Uncertainty-based Model Training for Index and Retrieval (UN-Model). UN-Index is responsible for constructing a high-confidence index, providing a high-quality index structure for matching. UN-Retrieval, building upon UN-Index, integrates relevance scores and uncertainty scores during the retrieval process to balance the relevance and novelty of matching results. UN-Model provides modeling capabilities for estimating relevance scores and uncertainty scores for both UN-Index and UN-Retrieval.

The main contributions of this paper are as follows:
\begin{itemize}

\item To improve the balance of relevance and novelty, this paper introduces a matching framework incorporating uncertainty, consisting of uncertainty-based index construction, uncertainty-based retrieval, and uncertainty-based modeling.

\item To the best of our knowledge, this paper proposes introducing uncertainty information during the index construction phase of the matching process to build a high-quality index for the first time.

\item On the deep matching direction based on indexing, UICR demonstrates a commendable performance in balancing relevance and novelty in recommendation systems.
\end{itemize}

\section{Methodology}

In this section, we first provide the problem definition in Sec. \ref{sec:Problem Definition}. Then, we introduce the pipeline for Uncertainty-based Index Construction and Retrieval(UICR) in Sec. \ref{sec:Uncertainty-based matching Pipeline}. Next, we present the uncertainty-based index(UN-Index) construction method in Sec. \ref{sec:Uncertainty-based Index Construction}. Then, we elaborate on the uncertainty-based retrieval(UN-Retrieval) algorithm in Sec. \ref{sec:Uncertainty-based Retrieval Methods}. Afterwards, we introduce the uncertainty estimate module, which will be utilized for user-to-item and item-to-item uncertainty estimation, in Sec. \ref{sec:Uncertainty Modeling}. Finally, we describe the uncertainty-based model training method in Sec. \ref{ref:Model Training}.

\subsection{Problem Definition}
\label{sec:Problem Definition}

The objective of matching phase in the Recommendation System is to retrieval a subset of items from the total candidate items, which may reach billion-scale. In this paper, we utilize $\mathcal{I}$ and $\mathcal{U}$ denote the item candidate set and user set respectively, $i$ and $u$ stand for a specific item $i \in \mathcal{I}$ and user $u \in \mathcal{U}$. The matching model can be defined as $\mathcal{M}(x_{u}, x_{i})$. It is a similarity mapping function according to some performance metrics. 

The index in matching stage is defined as $\mathcal{H}_{\mathcal{I}}$, which incorporates features of items in $\mathcal{I}$ and target to increace the retrieval efficiency of retrieval.
\vspace{-4pt} 
\begin{small}
\begin{equation}
    \label{eq:matching problem}
    \mathcal{B}(u)=argTopK_{i \in \mathcal{I}} \mathcal{F}(\mathcal{M}(x_{u}, x_{i}), \mathcal{H}_{\mathcal{I}})
\end{equation}
\end{small}
where $\mathcal{F}$ is the retrieval method based on matching model $\mathcal{M}$ and index $\mathcal{H}$, $\mathcal{B}(u)$ is the retrieval outcomes for user $u$, $x_{u}$ and $x_{i}$ are user and item features respectively. 

\subsection{Uncertainty-based Index Construction and Retrieval Pipeline}
\label{sec:Uncertainty-based matching Pipeline}

\setlength{\abovecaptionskip}{2pt} 
\begin{figure*}
    \centering
    \includegraphics[width=0.8\linewidth]{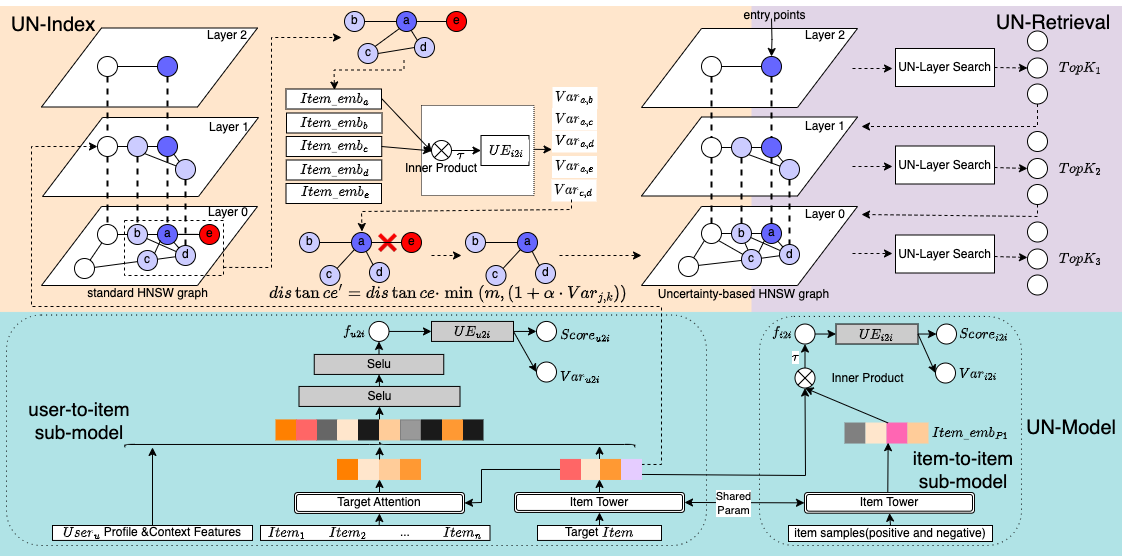}
    \caption{Uncertainty-based Matching Modeling Pipeline}
    \label{fig:UICR pipeline}
    \vspace{-15pt} 
\end{figure*}

In this paragraph, we describe the pipeline for constructing a uncertainty-based matching model. The pipeline consists of three parts: \textbf{UN-Index}(an item-to-item uncertainty-based index construction algorithm),\textbf{UN-Retrieval}(an user-to-item uncertainty-based beam search retrieval algorithm) and \textbf{UN-Model}(a multi-task model training method that considers uncertainty of both user-to-item and item-to-item).

In order to simultaneously enhance relevance and novelty during the matching phase, the design concept of UICR involves introducing uncertainty to improve the matching results with high estimated scores and low uncertainty, thus enhancing the relevance of matching results. By increasing percentage the matching results with high uncertainty, it improves the novelty of the results. Specifically, UN-Index focuses on improving relevance. During the construction of UN-Index, only index paths with high estimated scores and low uncertainty are retained, while paths with high uncertainty are removed. UN-Retrieval emphasizes enhancing novelty. During the retrieval process, a measurement scheme based on uncertainty is used to select top $k$ for users. UN-Model provides corresponding uncertainty and score estimation capabilities for both aspects.

The workflow of the UICR is as follow:
\begin{itemize}
\item \textbf{Step 1} The UN-Model is trained to develop the ability to estimate user-item relevance, item-item similarities, and uncertainty estimation. This enables us to predict the relevance of <user, item> pairs and estimate the similarity between items.

\item \textbf{Step 2} In the UN-Index module, the confidence index is constructed. Based on the item-item similarity and confidence, a high-confidence index is built.

\item \textbf{Step 3} In the UN-Retrieval module, using the index structure generated by the UN-Index module and the <user, item> model provided by the UN-Model module, a multi-layered search is performed to select the top-K recommendations.
\end{itemize}

\subsection{Uncertainty-based Index Construction Algorithm}
\label{sec:Uncertainty-based Index Construction}

\subsubsection{Why do we need to model the uncertainty in index construction?}

A significant practical issue in the industry is that matching model often faces candidate item sets ranging from millions to billions. Therefore, a common solution in the industry is to build index structures to boost the model retrieval efficiency. As we know, HNSW index is one of the most widely used method in productive environments. During the construction of the HNSW index, only the distances between item-to-item embeddings are considered to select the neighboring items for each item.

However, in practical industrial systems, due to various reasons, the estimation of distances is often not very accurate. The construction of the index relies on the distances between <item, item>, which can lead to bias in the resulting index. Consequently, the constructed index may not be the optimal one, thereby reducing the accuracy of the beam search results during the matching stage. There are many reasons for inaccurate distance estimation, such as the imbalance in item exposure caused by the Matthew effect. This imbalance leads to insufficient embedding learning for items other than those in the head, resulting in certain uncertainties in distance calculation.

Therefore, in the index construction phase, we introduce uncertainty to enhance each item's ability to select its truly close neighboring items, aiming to improve the accuracy of index construction. Considering the uncertainty in the index and comparing with the definition in equation (\ref{eq:matching problem}), matching model can be re-formulated as:
\begin{small}
\vspace{-1.2em}
\begin{equation}
    \label{eq:matching problem with optimized index}
    \mathcal{B}(u)=argTopK_{i \in \mathcal{I}} \mathcal{F}(\mathcal{M}(x_{u}, x_{i}), \mathcal{H}_{\mathcal{I}}^{'})
\end{equation}
\end{small}
\vspace{-0.8em}
where $\mathcal{H}_{\mathcal{I}}^{'}$ is the index constructed by UN-Index.

\subsubsection{Uncertainty-based Index Construction}
In this section, we introduce the uncertainty-based HNSW index construction method. The standard HNSW index construction process does not take the uncertainty of distance of each item-to-item pair into the consideration, thus, the candidates of subsequent beam search retrieval can be improved by incorporate both the uncertainty-based modeling and the uncertainty-based distance weighted calculation. The fundamental reason for choosing the uncertainty-based distance weighted calculation in index construction is to decrease the uncertainty of the distance between the points in the candidate set and the seed points during the search process of the beam search algorithm. Meanwhile, due to the decrease of uncertainty in the distance measurement of item-to-item pairs among the index, this contributes to enhancing the relevance of retrieval results. 

The calculation method of weighted distance between two item embeddings can be described as follow:
\begin{small}
\vspace{-0.5em}
\begin{equation}
    \label{eq:distance_weight}
    distance_{j,k}^{'} = distance_{j,k} \cdot min(m, (1 + \alpha \cdot var_{j,k})) 
\end{equation}
\vspace{-0.5em}
\end{small}
where $distance_{j,k}^{'}$ is the weighted distance between item $j$ and item $k$, $L_2$ $distance_{j,k}$ is the raw distance between item $j$ and item $k$ in HNSW index, $var_{j,k}$ is the variance of similarity between the embedding of item $j$ and item $k$, $m$ and $\alpha$ are hyper parameters need to be tuned. Hyper parameter $\alpha$ is a crucial parameter during the process of optimizing the constructed HNSW index and it should be set to larger than 0. According to equation (\ref{eq:distance_weight}), items with larger variances will have larger weighted distance values after weighting, while items with smaller variances will have smaller weighted distance values. The criterion for selecting which item will be retain is that the smaller the weighted distance, the nearer the item. In other words, items with closer weighted distances will be retained.

So the uncertainty-based index construction is implemented as follows,
\begin{itemize}
\item \textbf{Step 1} The standard HNSW index construction algorithm is used to build the original HNSW index. During the index construction process, the number of neighbours need to be saved of each point is set to $n$.
\item \textbf{Step 2} The item-to-item confidence module in the trained model is invoked to score each item-to-item pair and save the variance of each pair that are loaded from the HNSW index constructed in Step 1. 
\item \textbf{Step 3} We use the variance to weight distance between each item-to-item pair by equation (\ref{eq:distance_weight}) and the nearest $n^{'}$ items, where $n^{'} < n$, in the HNSW index built in Step 1 are retained.
\end{itemize}

Thus, we will ultimately utilize the uncertainty-based index with a neighbor count of $n^{'}$ for retrieval.

\vspace{-6pt} 
\subsection{Uncertainty-based Retrieval Algorithm}
\vspace{-4pt} 
\label{sec:Uncertainty-based Retrieval Methods}

\subsubsection{Why do we need to consider the uncertainty in retrieval process}

As shown in the Figure \ref{fig:four_quadrant}, items in the candidate set can be divided into four quadrants(The higher the variance (var), the greater the uncertainty). Items with high relevance and low uncertainty(low variance) are typically preferred for display, while items with low relevance and low uncertainty may not be prioritized for display. The remaining two quadrants, which consist of items with low relevance and high uncertainty, as well as items with high relevance and high uncertainty, may benefit from appropriate exposure opportunities to enhance the system's estimation capability for their relevance and novelty. Therefore, in the retrieval process, we need to take into account both relevance and uncertainty to determine whether to return items from the candidate set to the downstream pipeline of the recommendation system.

We discuss three scenarios: 1) In a matching model that does not consider variances, the items selected by the matching model are located in the first and fourth quadrants, representing highly correlated items. 2) When we retain items with low variances of distance among neighbors and do not consider the user-to-item variance of the matching model, compared to the scenarios 1), the number of items in the fourth quadrant increases, while the number of items in the other three quadrants decreases. Therefore, from the perspective of relevance, the matching performance of the model improves. 3) In UICR, when the index retains item-to-item neighbors with low variances and the model retains items with high user-to-item variances, the number of items in the fourth quadrant can be slightly increased by adjusting the parameters that control the impact of variance on the weight scores. At the same time, the number of items retained in the first and second quadrants is increased, allowing the model to improve performance in terms of both correlation and novelty dimensions.

\vspace{-0.5em}
\subsubsection{Uncertainty-based beam search Algorithm}
\vspace{-1em}
\begin{algorithm}
\small
\setlength{\belowcaptionskip}{-0.5cm}
        \setlength{\baselineskip}{8pt}
	\renewcommand{\algorithmicrequire}{\textbf{Input:}}
	\renewcommand{\algorithmicensure}{\textbf{Output:}}
	\caption{UN Layer Search}
	\label{alg:beam search}
	\begin{algorithmic}[1]
            \REQUIRE
            $ep$ is the entry points \\
            $l_{c}$ is the current layer number \\
            $T_{c}$ is the number of steps to search in the current layer \\
            \STATE $S \leftarrow ep$  \slash\slash  set of visited points \\
            \STATE $C \leftarrow ep$  \slash\slash  set for candidates \\
            \STATE $W \leftarrow ep$  \slash\slash  dynamic list of results \\
            \FOR{$t = 1 \to T_{c}$}
                \STATE $N \leftarrow$ union of neighbors at layer $l_{c}$ of all items in $C$
                \STATE $N \leftarrow N - S$ \slash\slash pruning visited items
                \STATE $S \leftarrow S \cup N$ \slash\slash mark as visited
                \STATE $W \leftarrow argTopK_{v \in W \cup U}S_{fusion}(u, \textbf{e}_{v})$
                \STATE $C \leftarrow W \cap N$ \slash\slash new candidates
            
            \ENDFOR
            \RETURN $W$

            \ENSURE $ef_{c}$ results with largest score s(u, v) to user $u$
	\end{algorithmic}  
\end{algorithm}
\vspace{-1em}

We enhance the original HNSW search algorithm by incorporating beam search with variance weighted calculation and propose a method called uncertainty-based beam search retrieval.

The search process of HNSW\cite{malkov2018efficient} entails navigating through a hierarchy of proximity graphs in a layer-wise and top-down fashion. Due to the utilization of the uncertainty-based distance weighted calculation approach during index construction, higher confidence and closer distance points are selected as neighbors, resulting in an improvement in the similarity of adjacent items during the index construction process. Consequently, the relevance of the items returned by the matching model is enhanced compared with considering only the relevance score of the user-to-item sub-model. Thus, by incorporating uncertainty-based score weighted approach, we can merge the model's confidence scores for user-item pairs with the relevance scores as follow:
\vspace{-0.5em}
\begin{small}
\begin{equation}
    \label{eq:fusion-score}
    S_{fusion}=score_{u2i}+\beta \cdot var_{u2i}
\end{equation}
\end{small}
\vspace{-0.7em}
where $\beta$ is a hyper-parameter needs to be tuned.

\begingroup
\begin{figure}[ht] 
    \centering
    \includegraphics[width=0.4\linewidth]{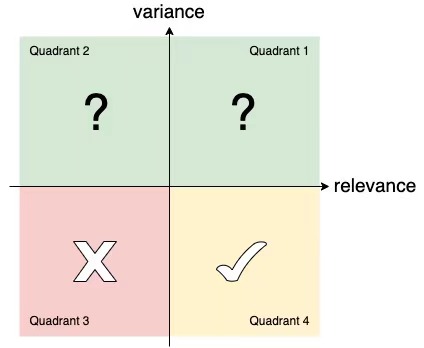}
    \caption{Generally, for a user, all items in the candidate set can be divided into four quadrants based on their objectively real but numerically unknown relevances and variances.}
    \label{fig:four_quadrant}
    \vspace{-2.2em} 
\end{figure}
\endgroup

\subsection{Uncertainty Estimate Module}
\label{sec:Uncertainty Modeling}

An ideal uncertainty modeling module in UICR should have the capability of estimating the variance with high accuracy for each relevance score with minimal online time and space complexity, while ensuring no loss in relevance.

In terms of implementation details, the input of this module is the logit function, noted as $f$, which can be formulated as follow:

\vspace{-1em}
\begin{small}
    \begin{equation}
        \label{eq:source-to-target-logit}
        f(x_{s}, x_{t})=\sigma^{-1}(\mathcal{M}_{s2t}(x_{s}, x_{t}))
    \end{equation}
\end{small}
\vspace{-1em}

where $\mathcal{M}_{s2t}$ is the sub-model of source-to-target, $x_{s}$ and $x_t$ are the source and target features.
The module's output consists of two values: the estimated score and the variance. The uncertainty estimation module of source-to-target part can be represented by the following formula:
\vspace{-0.4em}
\begin{small}
\begin{equation}
    \label{eq:source-to-target-uncertainty}
    <score_{s2t}, var_{s2t}>=UE_{s2t}(f(x_{s}, x_{t}))
\end{equation}
\end{small} 
\vspace{-0.3em}
where $UE_{s2t}$ is the Uncertainty Estimate module for source-to-target pair, $score_{s2t}$ and $var_{s2t}$ are the estimated relevance score and the corresponding variance. The loss function of the $UE_{s2t}$ module is noted as $\mathcal{L}_{s2t}$. In the specific tasks afterwards, this module will be utilized to estimate user-to-item and item-to-item pairs.We will introduce the training method of the $UE_{s2t}$ model in Sec. 2.6.

\subsection{Model Training}
\label{ref:Model Training}
\subsubsection{Uncertainty modeling algorithm}
According to the uncertainty-based index construction and beam search methods introduced in Sec. \ref{sec:Uncertainty-based Index Construction} and Sec. \ref{sec:Uncertainty-based Retrieval Methods}, it is obviously that the uncertainty modeling module required by the UICR algorithm needs to satisfy the following requirements: 1) The uncertainty modeling module needs to have low time and space complexity ,because the uncertainty-based beam search for user-to-item needs to be deployed as online services for the recommendation system. 2) The uncertainty modeling module needs to have strong model architectural compatibility, because in real productive environment, the matching model includes complex model structures, such as long sequence model structures, multi-objective model structures, multi-interest model structures, etc. 3) The uncertainty modeling module should be independent, capable of separately modeling the uncertainty of user-to-item and item-to-item, and each module should be able to work independently in inference calculations and uncertainty estimation. 

In the UICR framework, Deep Uncertainty-Aware Learning \cite{du2021exploration} is the method we choose to model the user-to-item and item-to-item uncertainty. The UICR framework does not require any specific method here for the uncertainty modeling. Therefore, other potential uncertainty model methods can also be used here. DUAL is based on Gaussian processes and it aims to provide predictive uncertainty estimations. At the same time, DUAL can be adapted widely in current industrial recommendation models by maintaining the flexibility of deep neural networks and no extra requirement of the architecture of the deep neural networks. DUAL can be easily implemented on existing models and deployed in real-time industrial recommendation systems.

The loss function of DUAL take both the relevance and the uncertainty into the consideration simultaneously, thus can be written as 
\begin{small}
\begin{equation}
    \label{eq:dual_loss_s2t}
    \mathcal{L}_{UE_{s2t}}=\mathcal{D}(f_{s2t}, y_{s2t})
\end{equation}
\end{small} 
where $\mathcal{L}_{UE_{s2t}}$ is the loss of $UE_{s2t}$ module for source-to-target pair, noted as $s$ and $t$ respectively, $\mathcal{D}$ is the DUAL module, $f_{s2t}$ is the logit function of similarity between source and target, $y_{s2t}$ is the label that whether source and target are similar.
\subsubsection{Modeling Uncertainty of Item-to-Item}

For the preparation of item-to-item similarity of training samples, UICR employs the item-to-item matching algorithm to select $Sample_{pos}$ items from the whole candidate item set with high similarity as positive samples. There are no special requirements in theory for selecting a specific item-to-item matching algorithm to train the $UE_{i2i}$ module of the UICR model. Classical item-to-item matching algorithms, like item-based Collaborative filtering, Swing \cite{yang2020large}, Enhanced Graph Embedding with Side Information (EGES) \cite{wang2018billion}, Node2vec\cite{grover2016node2vec}, can be used for this purpose. For simplicity, we have chosen the Swing algorithm, which is a widely used and classical item-to-item matching algorithm in the e-commerce domain. Within each training batch, UICR randomly sample $Sample_{neg}$ items from other instances as negative samples, forming the positive and negative sample sets for the items in the current instance. For each instance, UICR query the original feature data of the samples and use the item tower to perform forward inference, obtaining the corresponding item embeddings. Subsequently, UICR calculates the inner product between the current item embedding and the positive, negative samples respectively, multiply the inner product by the temperature coefficient, then feed the model's output to the $UE_{i2i}$ network. The loss function of the $UE_{i2i}$ network is noted as $\mathcal{L}_{i2i}$.

According to the method described above, the item-to-item similarity and variance can be calculated as follow:
\vspace{-0.4em}
\begin{small}
\begin{equation}
    \label{eq-item-similarity-variance}
    <score_{j,k}, var_{j,k}>=UE_{i2i}(\tau\textbf{e}_{j}^{T}\textbf{e}_{k})
\end{equation}
\end{small}
\vspace{-0.3em}
where $score_{j,k}$ is the similarity score between item $j$ and item $k$, $var_{j,k}$ is the variance of similarity score between item $j$ and item $k$, $\tau$ is the temperature coefficient.

\subsubsection{Modeling Uncertainty of User-to-Item}
Sample preparation for user-to-item uncertainty modeling is much easier. Because the training samples have positive and negative samples generated from multiple possible methods, like user click positive samples, negative sampling samples sampled by different sampling strategies. Generally speaking, the user-to-item uncertainty module can be formalized in equation (\ref{eq:user-to-item-uncertainty}). The output loss of the network is $\mathcal{L}_{u2i}$. Due to there exists no interaction between item tower input feature and user part of network. Equation (\ref{eq:user-to-item-uncertainty}) can be re-written as 
\vspace{-0.4em}
\begin{small}
\begin{equation}
    \label{eq:user-to-item-uncertainty}
    <score_{u_{j},k}, var_{u_{j},k}>=UE_{u2i}(NN_{user}(x_{c}, x_{u_{j}}, \textbf{e}_k))
\end{equation}
\end{small}
\vspace{-1.5em}

\subsubsection{Model Training}

In summary, we have completed the introduction to the various components of the model. The loss function for model training can be represented by the following equation,
\vspace{-0.4em}
\begin{small}
\begin{equation}
    \label{eq:total_loss}
    \mathcal{L}=\mathcal{D}(f_{u2i}, y_{u2i}) + \lambda_{i2i}\mathcal{D}(f_{i2i}, y_{i2i})
\end{equation}
\end{small}
\vspace{-0.4em}
where $\lambda_{i2i}$ is a hyper parameter which balance the user-to-item and item-to-item losses in the model optimization process.
The UICR model is trained using the Adam optimizer. It should be noted that due to the presence of the item-to-item uncertainty estimation network, there may be issues of data leakage and data crossing at multiple stages, including the item-to-item samples preparation in each train batch, the feature queried for the item tower and the order in which training data enters the network. It is crucial to handle these issues carefully to avoid data leakage and data crossing.

\subsubsection{Online Deployment}
The engineering team use Tensor-RT as the inference engine and enable various optimizations, including but not limited to multi-streaming, precomputing item embeddings offline and making them directly accessed online, FP $32$ to $16$. With the above optimizations, a complete UICR retrieval process takes approximately an average of $25ms$. In terms of machine performance, currently, we use GPU (A30), which can achieve several hundreds QPS per GPU.

\section{Experiments}

In this section, extensive offline and online experiments are performed on both the large-scale recommender system in Shopee and public benchmark datasets to answer the following research questions: 

\begin{itemize}
\item RQ1 Does our proposed method outperform the baseline methods?
\item RQ2 How does each part of our UICR model work? 
\item RQ3 How does the model perform when deployed online? 
\end{itemize}

Before presenting the evaluation results, we first introduce the experimental setup, including datasets, baselines, metrics, and parameter settings.

\subsection{Experimental Setup}

\subsubsection{Datasets}
We adopt public datasets and industrial datasets to comprehensively compare UICR models and baseline models. The statistics of the datasets are shown in Table \ref{tab:dataset}. 
\vspace{-6pt} 
\begin{table}[h]
  \caption{\textcolor{black}{Statistics of public and industrial datasets.}}
  \label{tab:dataset}
  \centering
  \setlength{\abovecaptionskip}{0.cm}
  \setlength{\belowcaptionskip}{-0.cm}
  \vspace{-4pt} 
  \begin{tabular}{c|cccc}
    \toprule
    Datasets & Users & Items & Samples & Density\\
    \midrule
    A      & \num{2.1e7} & \num{9.4e6} & \num{8.3e7} & \num{4.2e-7} \\
    T      & \num{9.8e5} & \num{4.2e6} & \num{1.0e8} & \num{2.4e-5} \\
    Shopee & \num{4.6e6} & \num{6.3e5} & \num{8.5e7} & \num{2.9e-5} \\
  \bottomrule
\end{tabular}
  \vspace{-1.5em}  
\end{table}

\textbf{A dataset\footnote{https://tianchi.aliyun.com/dataset/649}}: To validate the real-world effectiveness on large-scale industrial datasets, we utilize the complete dataset from Amazon, encompassing data from all categories including Books, Movies \& TV, Clothing, and others, while other papers that select only a few pure categories. We generate multiple samples by sorting the historical rating sequences of users by time. In other words, we use ${b_1...b_{n-1}}$ as user features and ${b_n}$ as the target item.

\textbf{T dataset\footnote{https://cseweb.ucsd.edu/\textasciitilde{}jmcauley/datasets.html\#amazon\_reviews}}: This dataset was first released by Alibaba Taobao and is widely used as a common benchmark in collaborative filtering approaches. It is a user behavior log for Taobao, including click, purchase, add-to-cart, and favorite behaviors.

\textbf{Industrial Shopee dataset}: This dataset is an industrial dataset collected by Shopee App which is one of the top-tier mobile Apps in Southeast Asia. Shopee is also a typical e-commerce scenario, with similar data density between the two.

\subsubsection{Baselines}

In the industry, the mainstream development paths for matching algorithms can be divided into three categories: 1) collaborative filtering based item-to-item algorithms, 2) vector model based retrieval algorithms, and 3) full-database retrieval algorithms based on complex DNN models (such as NANN\cite{chen2022approximate}). From a perspective of technological advancement, it is generally believed that 3) > 2) > 1). Therefore, we selected the recently industry iteration achievement NANN as our main baseline. For classical vector-based retrieval algorithms, we chose the classic YoutubeDNN and DSSM. Recent iterations have been based on them for multi-interest, multi-objective, and long-term interest iterations. However, these algorithm iteration achievements are extensions based on vector matching models. Because they can not introduce complex intersection architecture, like multi-head attention, to model user historical behavior sequence and candidate item.

We compare the UICR with classical matching methods and recent proposed methods as follows:
\begin{itemize}
\item \textbf{DSSM} DSSM\cite{huang2013learning} is originated from Natural Language Processing and it is one of the most widely used deep matching model in industial recommendation system.
\item \textbf{YoutubeDNN} YoutubeDNN\cite{covington2016deep} is one of the most successful deep learning based matching model widely used in industrial recommendation system.
\item \textbf{NANN} NANN upgrades the vector dot product during the retrieval process to a DNN model. This surpasses the upper limit of expressive power in vector models, resulting in significant improvements in the capabilities of the model.
\end{itemize}

\subsubsection{Metrics}
In our experiments, we use three metrics to evaluate the performance of the UICR model: $Recall@N$, $CateEntropy@N$ and $NewCateRatio@N$ \cite{vargas2011rank}. The first metric primarily assess the relevance of the matching results, while the latter two metrics primarily evaluate the novelty of the matching results.The calculation methods are as follows:
\begin{small}
\vspace{-0.5em}
\begin{equation}
    \label{eq:recall_at_N}
    \text{Recall@N} = {|P_u \cap G_u|}/{|G_u|} 
\end{equation}
\end{small}
where $P_u$($|P_u|=N$) denotes the set of retrieved items and $G_u$ denotes the set of ground truths. The higher the $Recall@N$, the better the relevance of the model's matching results.

The metric $NewCateRatio$ is defined as follow:
\begin{small}
\vspace{-0.5em}
\begin{equation}
    \label{eq:new-category-ration}
    \text{NewCateRatio@N}={|P_{u}\_Cate \setminus H_{u}\_Cate|}/{|H_{u}\_Cate|}
\end{equation}
\end{small}
where $P_{u}\_Cate$ stands for the set of categories of items in the retrieved top $N$ result, $H_{u}\_Cate$ means the set of categories of items in the user's historical behavior list, like user click sequence. Therefore, the larger the result computed by equation (\ref{eq:new-category-ration}), the better the novelty of the sequences returned by the matching model.

The metric $CateEntropy$ can be represented as follow,
\begin{small}
\vspace{-0.5em}
\begin{equation}
    \label{eq:CateEntropy}
    CateEntropy = - \sum_{i=1}^{K} Prob_{P_{u}\_Cate_{i}} \cdot \log_2(Prob_{P_{u}\_Cate_{i}})
\end{equation}
\end{small}
where $Prob_{P_{u}\_Cate_{i}}$ represents the probability distribution calculated by category in the matching results for user u. Equation(\ref{eq:CateEntropy}) computes the entropy distribution at the category level. If the entropy is larger, it indicates better diversity in the matching results.

\subsubsection{Parameter Settings}

We set the architecture of fully connected item tower in UICR to $[64, 48, 32]$ and the dimension of each feature to $32$. The final number of neighbors used in the HNSW index structure employed in this paper is uniformly set to 32.The fully connected layer in UICR, which take the context feature, target attention of user sequence and item candidate and item tower output embedding as input, is set to $[128, 64, 1]$. The temperature coefficient is set to $10.0$ by grid search. All the models reported in this paper are implemented in TensorFlow and are trained by Adam optimizer. The parameter $\lambda_{i2i}$ in equation (\ref{eq:total_loss}) is set to $0.1$.

\begin{table}
\setlength{\abovecaptionskip}{0cm}
\setlength{\belowcaptionskip}{0cm}
    \center
    \setlength\tabcolsep{1.0pt}
  \caption{Results of different methods on public and industrial datasets.All the results listed in this table are calculated with $100$ matching outcomes.}  
  \label{tab:all result}
  \begin{threeparttable}
  \small
  \begin{tabular}{c|c|ccc}
  \toprule[1.2pt]
    {DataSet} & {Algorithm} & Recall & Entropy & NewCateRatio \\
    \hline
    \multirow{6}*{\makecell{A}} & DSSM & 0.0185 & 1.9021 & 0.8538 \\
    ~ & YoutubeDNN & 0.0231 & 2.0482 & 0.8733 \\
    ~ & NANN & 0.0290 & 1.7261 & 0.8313 \\
    ~ & UICR & \textbf{0.0305} & 1.7583 & 0.8427 \\
    \hline
    \multirow{3}*{\makecell{T}} & DSSM & 0.0426 & 3.461 & 0.8195 \\
    ~ & YoutubeDNN & 0.0477 & 3.2613 & 0.8001 \\
    ~ & NANN & 0.0628 & 2.6481 & 0.7556 \\
    ~ & UICR & \textbf{0.0726} & 2.6693 & 0.7698  \\
    \hline
    \multirow{3}*{\makecell{Industrial}} & DSSM & 0.0864 & 1.6716 & 0.5377 \\
    ~ & YoutubeDNN & 0.0959 & 1.7298 & 0.5612 \\
    ~ & NANN & 0.1202 & 1.4482 & 0.4837 \\
    ~ & UICR & \textbf{0.1394} & 1.4865 & 0.4850 \\
  \bottomrule[1.2pt]
\end{tabular}
\end{threeparttable}
\vspace{-19pt}
\end{table}

\vspace{-4pt}
\subsection{RQ1: Does our proposed method outperform the baseline methods?}

In order to validate the effectiveness of the UICR algorithm, we selected three algorithms as baselines for comparison. The comparative performance of UICR and each baseline model on different datasets is reported in Table \ref{tab:all result}. We have the following observations: 1) DNN vs Vector: The user-to-item networks of NANN and UICR are both based on complex DNNs, while YoutubeDNN and DSSM are based on vector-based single/dual networks. In terms of expressive capability, DNNs are significantly superior to vectors. The $Recall@100$ data from the Table \ref{tab:all result} also shows that NANN/UICR significantly outperform YoutubeDNN and DSSM. 2) NANN vs UICR: In terms of metrics, UICR achieves improvements in both relevance and novelty metrics. This is attributed to the UN-Index and UN-Retrieval structures, where the UN-Index structure contributes to the improvement in $Recall@N$, and UN-Retrieval enhances novelty (for specific impacts, please refer to the subsequent ablation analysis). 3) In balancing relevance and novelty in recommendation systems, we believe that it is necessary to enhance novelty metrics while ensuring a certain level of relevance. Relevance determines short-term retention of users in the App, while novelty affects long-term user engagement. The greatest contribution of UICR is its ability to simultaneously consider relevance and novelty, achieving an improvement in novelty without compromising relevance.

More importantly, in the historical improvements of matching algorithms, there might have been a tendency to overlook novelty while focusing on improving relevance. We observe the trends in the experimental results of the three models (DSSM, YouTubeDNN, NANN). The improvement in various metrics for each model may come at the expense of sacrificing novelty, which is not conducive to the long-term ecological development of recommendation system.

\subsection{RQ2: How does each part of our UICR model work?}

We explore the contributions of each module in the UICR to the final results through statistical analysis and ablation experiments. 

The analysis in this section is based on user behavior data from the Shopee production environment (the Shopee dataset). This approach is taken because conclusions drawn from analysis using real-world data in industrial settings have more reference value and practical significance. The UICR introduces two additional modules compared to the baseline model: the $UE_{u2i}$ and $UE_{i2i}$ modules. Sec. \ref{sec:Uncertainty Modeling} and Sec. \ref{ref:Model Training} provide an introduction to the approach, data metrics, discussion and analysis of the ablation experiments.

\begin{table}
\setlength{\abovecaptionskip}{0cm}
\setlength{\belowcaptionskip}{0cm}
    \setlength\tabcolsep{1.0pt}
  \caption{Results of ablation study. All the results listed in this table are calculated with $100$ matching outcomes.}  
  \label{table:ablation_study}
  \begin{threeparttable}
  \small
  \begin{tabular}{l|ccc}
  \toprule[1.2pt]
    Experiment Group & Recall & Entropy & NewCateRatio \\
    \hline
    A:UICR w/o $UE_{i2i}$ \& w/o $UE_{u2i}$ & 0.1202 & 1.4482 & 0.4837 \\
    B:UICR w/o $UE_{i2i}$ \& \space\space w \space $UE_{u2i}$ & 0.1247 & \textbf{1.4904} & \textbf{0.4851} \\
    C:UICR \space\space w \space $UE_{i2i}$ \& \space\space w \space $UE_{u2i}$(Full UICR) & \textbf{0.1394} & 1.4865 & 0.4850 \\
  \bottomrule[1.2pt]
\end{tabular}
\end{threeparttable}
\vspace{-2.0em}
\end{table}

\subsubsection{Effect of UN-Index: the item-to-item Uncertainty Estimation Module}

\begin{figure}[ht]
    \centering
    \includegraphics[width=0.9\linewidth]{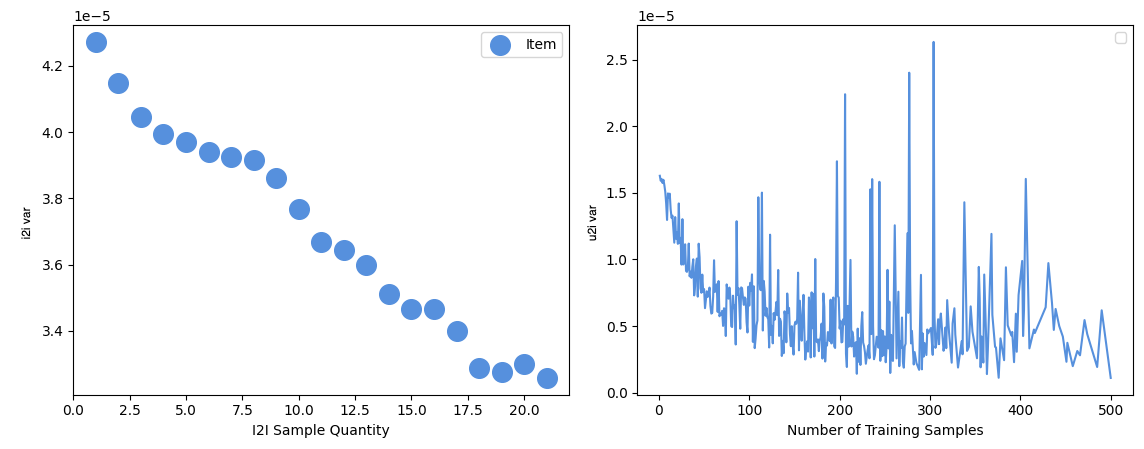}
    \vspace{-0.5em}
    \caption{Relationship between i2i/u2i uncertainty and number of training samples}
    \label{fig:i2i-confidence}
\end{figure}

A simple method to test the effectiveness of item-to-item confidence is to bucketize the click counts of items in the training dataset. The click counts are grouped into buckets, where each bucket represents a range of $10$ clicks. Then, $100$ items and their corresponding embeddings are randomly sampled without replacement from the items with click counts in the range $(10, 20]$. Similarly, $100$ items are randomly sampled without replacement from each of the other buckets. The $UE_{i2i}$ is used to estimate the variance between the different groups and the average value of variances is calculated. The results are shown in the left part of Figure \ref{fig:i2i-confidence}. We can observe that as the number of training samples increases, the average value of variances of the item-to-item gradually decreases, indicating an increase in confidence in the estimated scores. This suggests two things: 1) Our strategy for constructing item-to-item samples is effective. 2) The UN-Model is able to simultaneously model the item-to-item similarity score and the uncertainty.

\setlength{\intextsep}{1pt} 
\vspace{-1pt} 

We trained a version of the complete UICR model by removing the item-to-item uncertainty modeling module ($UE_{i2i}$). The performance of this model is presented in Table \ref{table:ablation_study} (Group \textbf{B} vs Group \textbf{C}). Based on the data, we can draw the conclusion that When the UN-Index module is removed (Group \textbf{B}), compared to Full UICR (Group \textbf{C}), the model's $Recall@100$ decreases, aligning with the training goal of UN-Index to enhance the relevance of the index.

\subsubsection{Effect of UN-Retrieval: the user-to-item Uncertainty Estimation Module}

Similar to the uncertainty validation method for item-to-item, the method for validating the effectiveness of user-to-item uncertainty involves grouping the click counts of each item in the training dataset. From each group, $10$ users are randomly sampled without replacement. The $UE_{u2i}$ network is invoked to estimate the uncertainty between the different groups, and the average variance of user-to-item is calculated. The results are shown in the right part of Figure \ref{fig:i2i-confidence}. We can observe that as the number of training samples increases, the variance of the user-to-item gradually decreases, indicating an increase in confidence in the estimated scores. This suggests that the model is able to simultaneously model both the estimated scores and their confidence.

In order to evaluate the user-to-item confidence modeling module. We also trained a version of the complete UICR model by removing the user-to-item confidence modeling component ($UE_{u2i}$). The performance of this model is presented in Table \ref{table:ablation_study} (Group \textbf{A} vs Group \textbf{B}). It is worth noting that when we introduced the UN-Retrieval, there was a slight increase in the Recall metric. It could be attributed to the addition of the uncertainty estimation module for user-to-item pairs (the Dual algorithm), which results in a slight increase in the number of parameters in the network model structure. Upon reviewing the Dual paper, we found that its inclusion also leads to a slight increase in AUC in CTR task, as the introduction of Dual brings about a slight increase in the model's performance due to the incorporation of additional parameters. As for the novelty-related metrics, there was indeed a significant improvement after the addition of UN-Retrieval, indicating exploration of the high-variance portion during the retrieval process.

\vspace{-1em}
\subsection{RQ3: How does the model perform when deployed online?}

\subsubsection{Online A/B experiment Result}

We deployed the UICR in the display advertising system on the Shopee App homepage. The online A/B test experiments showed that the UCIR algorithm resulted in an increase of $4.80\%$ in Revenue, an increase of $2.59\%$ in CTR.

\vspace{-1em}
\section{Related Works}
\textbf{Matching model development.} Deep matching algorithms are widely prevalent in the industry, personalized recommendation results are generated by constructing positive and negative samples along with network structures.
\cite{covington2016deep,huang2020embedding}
Moreover, research has revealed that employing a singular user embedding proves challenging in effectively capturing the entirety of a user's interests \cite{li2019multi,cen2020controllable,tan2021sparse} . Consequently, the paradigm of multi-embedding interest modeling approaches has emerged.

\textbf{Indexing Methods.} As the number of candidate items increases, dual-tower matching model encounters performance bottlenecks. Numerous research endeavors have leveraged tree \cite{zhu2018learning,zhu2019joint,zhuo2020learning} and graph structures \cite{chen2022approximate} for modeling purposes.
Deep retrieval (DR) [8] encodes all candidate items
with learnable paths and train the item paths along with the deep
model to maximize the same objective.

\vspace{-1em}
\section{Conclusion}
In summary, UICR method makes a significant contribution to the relevance and novelty in the matching field. This method includes UN-Index, UN-Retrieval and UN-Model, introducing uncertainty-related information during the matching stage to effectively address the challenge of balancing relevance and novelty. By incorporating uncertainty into the matching framework, it not only enhances the quality of the index (relevance) but also improves the novelty of recommendation systems. Additionally, our proposed method has been thoroughly validated through extensive experimental results.

\vspace{-1em}
\section{ACKNOWLEDGEMENTS}
We deeply appreciate Shujie Ma for his helpful suggestions and discussions.

\bibliographystyle{ACM-Reference-Format}
\bibliography{sample-base}

\appendix

\end{document}